\documentclass[12pt,a4]{article}

\usepackage{amsthm,amsmath,latexsym,amssymb,amsfonts,amscd}
\usepackage{graphics,lscape,fancyhdr,array,stmaryrd,euscript}
\usepackage{empheq}
\usepackage{dsfont}
\usepackage{verbatim}
\usepackage{color,tikz,tikz-cd}
\usetikzlibrary[snakes]
\usepackage{relsize,slashed}
\numberwithin{equation}{section}
\usepackage{hyperref}
\newcommand{\pl}{\partial}

\usepackage{hyperref}

\usepackage[numbers,sort&compress]{natbib}
\usepackage[nottoc]{tocbibind}

\newcommand{\aAb}{{\ensuremath{\boldsymbol{\mathcal{A}}}}}
\newcommand{\aBb}{{\ensuremath{\boldsymbol{\mathcal{B}}}}}

\newcommand{\ga}{\alpha}
\newcommand{\gb}{\beta}
\newcommand{\gc}{\gamma}
\newcommand{\gd}{\delta}

\newcommand{\fud}[2]{{}^{#1}{}_{#2}\,}
\newcommand{\fdu}[2]{{}_{#1}{}^{#2}\,}

\newcommand{\Pd}{{\mathtt{P}}}
\newcommand{\Ld}{{\mathtt{L}}}

\newcommand{\besubeqs}{\begin{subequations}}
\newcommand{\esubeqs}{\end{subequations}}

\newcommand{\dd}{{\mathrm{d}}}

\newcommand{\aomega}{{\underline{\omega}}}
\newcommand{\auxC}{{\underline{C}}}

\newcommand{\momega}{{\boldsymbol{\omega}}}
\newcommand{\mC}{{\boldsymbol{C}}}

\newcommand{\mg}{{\boldsymbol{g}}}

\newcommand{\Ainf}{\mathcal{A}}

\newcommand{\hs}{{\ensuremath{\mathrm{hs}}}}
\newcommand{\hsdouble}{{\ensuremath{{}^\Gamma \mathrm{hs}}}}
\newcommand{\ass}{{\mathrm{A}}}
\newcommand{\hsdeformed}{{\ensuremath{\boldsymbol{\mathrm{A}}_u}}}

\newcommand{\gers}[1]{{\llbracket #1\rrbracket}}

\begin{document}
%%%%%%%%%%%%%%%%%%%%%%%%%%%%%%%%%%%%%%%%%%%%%%%%%%%%%%%%%%
\pagenumbering{gobble}
\hfill
%\vspace{-1.5cm}
\begin{flushright}
    {}
\end{flushright}
\vskip 0.01\textheight
\begin{center}
{\Large\bfseries 
Formal Higher Spin Gravities\\
\vspace{0.4cm}}

\vskip 0.03\textheight

Alexey \textsc{Sharapov}${}^{1}$ and Evgeny \textsc{Skvortsov}${}^{2,3}$

\vskip 0.03\textheight

{\em ${}^{1}$Physics Faculty, Tomsk State University, \\Lenin ave. 36, Tomsk 634050, Russia}\\
\vspace*{5pt}
{\em ${}^{2}$ Albert Einstein Institute, \\
Am M\"{u}hlenberg 1, D-14476, Potsdam-Golm, Germany}\\
\vspace*{5pt}
{\em ${}^{3}$ Lebedev Institute of Physics, \\
Leninsky ave. 53, 119991 Moscow, Russia}

\end{center}

\vskip 0.02\textheight

\begin{abstract}
We present a complete solution to the problem of Formal Higher Spin Gravities --- formally consistent field equations that gauge a given higher spin algebra and describe free higher spin fields upon linearization. The problem is shown to be equivalent to constructing a certain deformation of the higher spin algebra as an associative algebra. Given this deformation, all interaction vertices are explicitly constructed. All formal solutions of the equations are explicitly described in terms of an auxiliary Lax pair, the deformation parameter playing the role of the spectral one. We also discuss a natural set of observables associated to such theories, including the holographic correlation functions. As an application, we give another form of the Type-B formal Higher Spin Gravity and discuss a number of systems in five dimensions.

\end{abstract}
\newpage
\tableofcontents
\newpage
%%%%%%%%%%%%%%%%%%%%%%%%%%%%%%%%%%%%%%%%%%%%%%%%%%%%%%%%%%
\section{Introduction}
%%%%%%%%%%%%%%%%%%%%%%%%%%%%%%%%%%%%%%%%%%%%%%%%%%%%%%%%%%
\pagenumbering{arabic}
\setcounter{page}{2}
Higher Spin Gravities describe dynamics of (usually) infinite multiplets that contain massless fields of arbitrarily high spin. Generally they are AdS/CFT dual of free CFT's \cite{Sundborg:2000wp,Sezgin:2002rt,Klebanov:2002ja}. Free CFT's enjoy infinite-dimensional global symmetries manifested in higher spin algebras. Interactions in Higher Spin Gravities are expected to be completely determined by gauging these higher spin algebras. We show that taking into account higher spin symmetry, but leaving aside locality, the problem of formally consistent and gauge invariant equations of motion can be solved in full generality. Constructing classical Higher Spin Gravity turns out to be equivalent to finding a one-parameter family of specific associative algebras that deform a given higher spin algebra. The latter observation drastically simplifies the problem. In particular, the equations can also be explicitly solved and a class of natural observables, including the holographic correlation functions, can be described. 

When understood as holographic duals of free CFT's, Higher Spin Gravities should be reconstructable from the free CFT correlation functions \cite{Heemskerk:2009pn,Koch:2010cy,Bekaert:2015tva,Sleight:2016dba}. The latter are completely fixed by global higher spin symmetries on the CFT side \cite{Maldacena:2011jn,Boulanger:2013zza,Alba:2013yda,Alba:2015upa}. Therefore, in the context of Higher Spin Gravities the only initial data is given by a higher spin algebra that is directly extracted from its free CFT dual (or from the $N\rightarrow\infty$ limit of an interacting one that is related to the free one via a double-trace deformation). The CFT origin makes it clear that higher spin algebras are associative algebras resulting from quotients of the universal enveloping algebra of the conformal algebra $U(so(d,2))$. Equivalently, higher spin algebras result from the deformation quantization of the algebra of functions on the coadjoint orbit associated with the free field as an irreducible representation of $so(d,2)$.

From the general point of view, a given higher spin algebra is just an associative algebra. The equations of motion of Formal Higher Spin Gravities look schematically like
\begin{align}\label{mastereq}
    d\Phi&=\mathcal{V}_2(\Phi,\Phi)+\mathcal{V}_3(\Phi,\Phi,\Phi)+\ldots =F(\Phi)\,,
\end{align}
where the bilinear structure map $\mathcal{V}_2(\bullet,\bullet)$ is completely determined by a given higher spin algebra. The equations look very similar to those of String Field Theory, see e.g. \cite{Gaberdiel:1997ia,Erler:2013xta} and, indeed, an appropriate mathematical framework to abstract gauging of higher spin symmetries is that of $L_\infty$-algebras, $Q$-manifolds and closely related ones. The reason for the qualifier {\it formal} is two-fold: first of all, there is a interesting relation to the formality theorems \cite{Sharapov:2017yde} and deformation quantization \cite{Sharapov:2018kjz}; secondly, due to locality not being imposed,\footnote{or even not being possible to impose since the interactions are known be singular \cite{Bekaert:2015tva,Sleight:2017pcz,Ponomarev:2017qab}.} usual field-theoretical computations may be problematic \cite{Giombi:2009wh,Giombi:2010vg,Boulanger:2015ova,Skvortsov:2015lja}, so the equations are formal in this sense too. It remains to be seen what kind of treatment higher spin theories are amenable to since they are not quite conventional field theories. 

A particular instance of the Formal Higher Spin Gravity problem was formulated in \cite{Vasiliev:1988sa}, with some specific solutions obtained in \cite{Vasiliev:1990vu,Prokushkin:1998bq,Vasiliev:2003ev}. We solve this problem in full generality by showing that: (i) the first-order deformations $\mathcal{V}_3$ in  \eqref{mastereq} are specified by the Hochschild cohomology of a given higher spin algebra, which is always nontrivial; (ii) as a result, there exists a one-parameter family $\hsdeformed$ of associative algebras that deform the higher spin algebra; (iii) the most important result is that all the vertices $\mathcal{V}_n$ can be expressed in terms of $\hsdeformed$; (iv) the equations are completely integrable and can be solved with the help of a Lax pair; (v) there is a number of invariants and covariants that are associated with the equations. In particular, there is a set of invariants that can compute the holographic correlation functions. More generally, given any one-parameter family of algebras we can construct and solve certain nonlinear gauge-invariant equations.    

These results should be contrasted with other formal solutions \cite{Vasiliev:1990vu,Prokushkin:1998bq,Vasiliev:2003ev}, aka Vasiliev equations. The advantages of our approach are full generality and simplicity: one can construct Formal Higher Spin Gravities directly, avoiding resolutions \cite{Sharapov:2017yde,Sharapov:2017lxr,Sharapov:2018hnl,Sharapov:2018ioy} which are highly ambiguous, cumbersome and may not capture the relevant observables. The new approach also clarifies the algebraic structures involved in the construction. As an application, we give another realization for the Type-B theory that was constructed by different means in \cite{Grigoriev:2018wrx}. Also, we discuss a number of systems in five dimensions with and without propagating degrees of freedom. 

The  paper is organized as follows. In Sect. \ref{sec:FHSG}, we formulate the problem of gauging higher spin symmetries and solve it by explicitly constructing the deformation, describing the solution space and observables. In Sect. \ref{sec:TypeB}, we work out an example of the Type-B Formal Higher Spin Gravity. In Sect. \ref{sec:FiveD}, we discuss two novel systems in $AdS_5$.

%%%%%%%%%%%%%%%%%%%%%%%%%%%%%%%%%%%%%%%%%%%%%%%%%%%%%%%%%%
\section{Formal Higher Spin Gravities}
\label{sec:FHSG}
%%%%%%%%%%%%%%%%%%%%%%%%%%%%%%%%%%%%%%%%%%%%%%%%%%%%%%%%%%
The plan is to briefly recall the definition of $Q$-manifolds, $L_\infty$-algebras and Free Differential Algebras (FDA) and how they can be used to formulate formally consistent equations. Next, we formulate problem of Formal Higher Spin Gravities, following essentially \cite{Vasiliev:1988sa}, but in full generality. Then, we solve it in a constructive way for any higher spin algebra, i.e., for any free CFT dual. We explicitly describe the solution space of the associated formal equations of motion by constructing the Lax pair.  Lastly, we list some invariants associated with the nonlinear equations. 

%%%%%%%%%%%%%%%%%%%%%%%%%%%%%%%%%%%%%%%%%%%%%%%%%%%%%%%%%%
\subsection{\texorpdfstring{$Q$}{Q}-manifolds, Free Differential Algebras and Formal Equations}
\label{sec:}
%%%%%%%%%%%%%%%%%%%%%%%%%%%%%%%%%%%%%%%%%%%%%%%%%%%%%%%%%%
There are three closely related objects (ranging them from more general to less general): $Q$-manifolds \cite{Alexandrov:1995kv}, $L_\infty$-algebras and Free Differential Algebras (FDA's) \cite{Sullivan77,Nieuwenhuizen:1982zf,DAuria:1980cmy}, see also \cite{Boulanger:2008up,Li:2018rnc}. $Q$-manifold is a super-manifold equipped with an odd, nilponent vector field $Q$ squaring to zero. The simplest example, is the $Q$-manifold associated with a Lie algebra: the coordinates $x^a$ on the Lie algebra are turned into odd ones $w^a$ and the odd vector field is $Q= l_{ab}^c\, w^a w^c \pl/\pl w^c$. The condition  $Q^2=0$
is then equivalent to the Jacoby identity for the structure constants $l_{ab}^c$ of the Lie algebra.

If $p$ is a stationary point of $Q$, that is $Q|_p=0$, then the Taylor expansion of $Q$ at $p$ gives an infinite collection of numerical coefficients: 
\begin{align}
    Q^a&=l^a_b\, w^b +l^a_{bc}\, w^b w^c+\ldots \,.
\end{align}
The integrability condition $Q^2=0$ leads then to a set of quadratic relations for the $l$'s that can be recognized as the definition of an $L_\infty$-algebra with the structure constants $\{l^a_{b_1\cdots b_n}\}_{n=1}^\infty$. It is important to note that $Q$-manifolds are globally defined, while every stationary point leads to an $L_\infty$-algebra, which is a local object in this sense. 

We define Free Differential Algebras as $L_\infty$-algebras whose underlying graded space has only non-positive degrees.\footnote{The restriction to non-positive degrees is chosen to match the one later in the text. It is well-known that the BV-BRST formulation of gauge theories and $Q$-manifolds, $L_\infty$-algebras are closely related to each other \cite{Barnich:2000zw}. FDA corresponds to some kind of minimal classical formulation with ghosts and anti-fields excluded, see \cite{Barnich:2004cr,Barnich:2010sw} for more detail.} Since FDA's is the main subject of the paper, let us unfold the definition in a bit more detail. Let $W^\aAb$ be the coordinates on some graded space $V$. The degree of $W^\aAb$ is denoted by $-|\aAb|$ and we assume that $|\aAb|\geq0$. Given some base manifold $\mathcal{M}$ with coordinates $x$, we can consider maps $\Omega^\bullet(\mathcal{M})\rightarrow V$ from the space of differential forms on $\mathcal{M}$ to $V$, which we also denote by $W^\aAb\equiv W^\aAb(x)$. Hence, $W^\aAb$ can be viewed as a differential form of degree $|\aAb|$ on $\mathcal{M}$. The associated {\it formal equations} are 
\begin{align}
\label{UnfldEquationsA} \dd W^\aAb = F^\aAb(W)\,,
\end{align}
where $\dd$ is the de Rham differential on $\mathcal{M}$, $\dd^2=0$. The odd vector field $F^\aAb(W)$ has the Taylor expansion of the form
\begin{align}
    F^\aAb(W)=\sum_{n}\sum_{|\aBb_1|+\cdots+|\aBb_n|=|\aAb|+1}
    f^\aAb_{\phantom{\aAb} \aBb_1\ldots \aBb_n}W^{\aBb_1}\wedge\ldots\wedge
    W^{\aBb_n}\,.
\end{align}
The associated vector field is $Q=F^\aAb \pl/\pl W^\aAb$. The nilpotency of $Q$ or the formal integrability of \eqref{UnfldEquationsA} coming from $\dd^2=0$ imply that  
\begin{align}
    \label{UnfldJacobi} F^\aBb\wedge\frac{\delta F^\aAb}{\delta
W^\aBb}\equiv0\,.
\end{align}
The formal equations are defined up to field-redefinitions, which correspond to different choices of coordinates in the target space $V$. The formal equations are also invariant under the gauge transformation
\begin{align}
    \label{UnfldGauge}\delta_{\epsilon} W^\aAb&=\dd\epsilon^\aAb+\epsilon^\aBb\pl_\aBb F^\aAb\,,
\end{align}
where the gauge parameter $\epsilon^\aAb$ has form degree $|\aAb|-1$. In particular, there is no $\epsilon^\aAb$ associated with $W^\aAb$ of degree zero, $|\aAb|=0$. 

The $Q$-manifolds associated with the simplest higher spin theories have two coordinates: odd, degree-one, $\omega$ and even, degree-zero, $C$, both taking values in a given higher spin algebra. It is important to stress that Eq. \eqref{UnfldEquationsA} may not be a well-defined PDE unless certain further constraints, e.g. locality, are imposed; hence, the qualifier {\it formal}. It is easy to come up with examples of equations that are troublesome as field theories/PDE's, see e.g. \cite{Vasiliev:1988sa,Vasiliev:1990vu,Boulanger:2015ova,Skvortsov:2015lja}. For dimension $d$ sufficiently small the equations turn out to be well-defined, e.g. for $d=1$ one is left with ODE's. There can be some critical $d^*$ (equal to the functional dimension of the target space) where the problem of well-posedness is subtle. For $d>d^*$ the equations become over-determined.

%%%%%%%%%%%%%%%%%%%%%%%%%%%%%%%%%%%%%%%%%%%%%%%%%%%%%%%%%%%%%
\subsection{Problem}
%%%%%%%%%%%%%%%%%%%%%%%%%%%%%%%%%%%%%%%%%%%%%%%%%%%%%%%%%%%%%
Let us now formulate the problem of Formal Higher Spin Gravities. It is useful to explain the ingredients from the AdS/CFT perspective. The problem is to write down formal equations \eqref{UnfldEquationsA} that would gauge the higher spin algebra associated with any given free CFT. We take some free field $\varphi$. The higher spin algebra is, roughly speaking, the algebra of all linear transformations of the space of the on-shell states of $\varphi$. Let us denote this space $S$, which is obviously an $so(d,2)$-module. Heuristically, the higher spin algebra $\hs$ is then\footnote{Dealing with such expressions may be subtle since $S$ is an infinite-dimensional vector space, see e.g. \cite{Basile:2018dzi}.} $S\otimes S^*$. It is infinite-dimensional and contains the conformal algebra $so(d,2)$ as a subalgebra. By construction, the algebra $\hs$ results from the universal enveloping algebra $U(so(d,2))$ upon factoring out certain two-sided ideal $I$ that annihilates $S$ (in the field-theoretical language, the ideal contains, e.g., the  wave-operator for $\varphi$). The field $\varphi$ being free, $\hs$ is not only a Lie algebra, but an associative one. The spectrum of single-trace operators $O_1\sim\varphi\pl\cdots \pl\varphi$ built out of $\varphi$ is given by $S\otimes S$. These operators are dual to the on-shell states of the $AdS$ dual. Here we assume that $\varphi$ takes values in the vectorial representation of some symmetry group, e.g. $U(N)$, that is then weakly gauged. Therefore, the only single-trace operators are bilinears, i.e.,  $S\otimes S$. Since $\varphi$ is free the set $O_1$ includes infinitely many conserved tensors, which are dual to massless fields in $AdS_{d+1}$. 

The higher spin algebra $\hs$ is a global symmetry on the CFT side and it should be gauged on the AdS side, hence, one-form connection $\omega$ of $\hs$ is an appropriate object. A special feature  of higher spin theories is that the bulk on-shell states are formally equivalent to the algebra $\hs$ itself, $S\otimes S \sim S\otimes S^*$, but we need to account for the map between $S$ and $S^*$ in doing so \cite{Iazeolla:2008ix}. This map is the inversion automorphism on the CFT side or, on the AdS side, is the automorphism that flips the sign of the AdS translations $\pi(P_a)=-P_a$ and leaves the Lorentz generators invariant, $\pi(L_{ab})=L_{ab}$.\footnote{The AdS algebra commutation relations are \begin{align*}
    [L_{ab},L_{cd}]&= L_{ad}\eta_{bc}+\text{3 more}\,, &[L_{ab},P_c]&=P_a\eta_{bc}-P_b\eta_{ac}\,, && [P_a,P_b]=L_{ab}\,.
\end{align*} 
It is obvious that $\pi$ is an automorphism. It extends to the automorphism of the higher spin algebra.} The degrees of freedom can be described by a zero-form $C$ with values in $\hs$ provided the action of the algebra on it is twisted by $\pi$. Given these data, the simplest system of equations reads 
\besubeqs\label{freesystem}
\begin{align}
\dd\omega&=\omega\star \omega\,,\\
\dd C&=\omega\star C-C\star \pi(\omega)\,.
\end{align}
\esubeqs
Here $\star$ denotes the product in $\hs$. The simplest exact solution is to take $\omega$ be a flat non-degenerate connection of the anti-de Sitter algebra $so(d,2)$. Such $\omega$ describes the empty $AdS_{d+1}$-space. Then, $C$ describes the right set of degrees of freedom. For the practical purposes, the elements of $\hs$ can be understood as functions $f(P_a,L_{ab})$ modulo some relations implied by the ideal $I$. The first few Taylor coefficients in
\besubeqs
\begin{align}
    \omega=\omega(P_a,L_{ab})&= A\, \mathrm{1}+e^aP_a +\varpi^{a,b}L_{ab}+\ldots\,,\\
    C=C(P_a,L_{ab})&=\phi \, \mathrm{1}+ F^{a,b}L_{ab}+W^{ab,cd}L_{ab}L_{cd}+\ldots\,,
\end{align}
\esubeqs
can be identified with the spin-one gauge potential $A$ (dual of the current $\varphi \pl \varphi$, if present), vielbein $e^a$, spin-connection $\varpi^{a,b}$, scalar $\phi$ (dual of $\varphi\varphi$), Maxwell field strength $F^{a,b}$, Weyl tensor $W^{ab,cd}$ and higher spin generalizations thereof.

Eqs. \eqref{freesystem} describe free fields. The central problem of Formal Higher Spin Gravity is to construct consistent deformations of the trivial system \eqref{freesystem}:
\besubeqs\label{problem}
\begin{align}
\dd\omega&=\omega\star \omega + \mathcal{V}_3(\omega,\omega,C)+\mathcal{V}_4(\omega,\omega,C,C)+O(C^3)=F^\omega(\omega,\omega|C)\,,\\
\dd C&=\omega\star C-C\star \pi(\omega)+\mathcal{V}_3(\omega,C,C)+O(C^3)=F^C(\omega|C)\,,
\end{align}
\esubeqs
where we simply added all possible terms based on the form-degree counting. Clearly, it is $C$ that is responsible for the non-triviality of the deformation.\footnote{Suppose that we have only $\omega$, then the only deformations one can add $\dd\omega=\omega\star \omega +\Psi(\omega,\omega)$ would correspond to the deformation of the higher spin algebra as a Lie algebra and usually there are none.} The odd vector field with components $F^\omega$, $F^C$ is, of course, required to be nilpotent, which is the only restriction on the vertices $\mathcal{V}$. The `expansion parameter' here is $C$ and in this sense we are trying to observe the higher spin $Q$-manifold from the  point $C=0$ (it is not actually a point as $\omega$ is not restricted to be small and it would be interesting to understand the global structure of this $Q$-manifold). Deviation from flatness is thus controlled by $C$, which is supposed to be small. Therefore, the problem is to relate the vertices $\mathcal{V}$ to the higher spin algebra $\hs$ that underlies the free equations \eqref{freesystem}.

%%%%%%%%%%%%%%%%%%%%%%%%%%%%%%%%%%%%%%%%%%%%%%%%%%%%%%%%%%%%%
\subsection{Solution}
%%%%%%%%%%%%%%%%%%%%%%%%%%%%%%%%%%%%%%%%%%%%%%%%%%%%%%%%%%%%%
The above problem of constructing the vertices of the Formal Higher Spin Gravity \eqref{problem} can be solved in full generality and the solution presented below is based on the ideas from (non-commutative) deformation quantization and slightly-broken higher spin symmetries studied in our recent paper \cite{Sharapov:2018kjz}. The applicability of our approach goes beyond the higher spin algebras as it is clear that \eqref{problem} requires very little data. Therefore, we assume that we are given some associative algebra $\hs$ and a finite group $\Gamma$ of its automorphisms (in our case $\Gamma=\mathbb{Z}_2=(1,\pi)$ with $\pi^2=1$).\footnote{The system \eqref{problem} seems to take the advantage of the Lie part of $\hs$ only (via commutators), but we should also keep in mind that it is easy to extend the dual free CFT to another one with global symmetries, say $u(M)$, which would correspond to gauging the Lie part of (appropriate real form of) $\hs\otimes gl(M)$. The latter encodes the full associative structure on $\hs$ for a sufficiently large $M$. Our solution will be applicable for $\hs\otimes gl(M)$ as well. \label{matrixcomment}} 

The main claim is that everything (vertices, solutions, invariants) can be obtained from a certain associative algebra constructed out of a given higher spin algebra. 
The first step is to realize \cite{Sharapov:2017yde,Sharapov:2018kjz} that a nontrivial deformation of \eqref{problem} is induced by 
\begin{align}\label{firstV}
    \mathcal{V}_3(\omega,\omega,C)&= \phi(\omega,\omega)\star \pi (C) \quad \text{or}\quad \pi (C) \star \phi(\omega,\omega)\,,
\end{align}
the two forms being equivalent and we stick to the first, i.e., the left form. Here, $\phi(a,b)$ is a nontrivial Hochschild two-cocycle with values in the adjoint representation twisted by $\pi$:
\begin{align}\label{twistedcocycle}
   (\delta \phi)(a,b,c)\equiv a\star \phi(b,c)-\phi(a \star b,c)+\phi(a,b\star c) -\phi(a,b)\star \pi(c)&=0\,.
\end{align}
It must be a nontrivial $\delta$-cocycle for the $\phi$ to induce $\mathcal{V}$ that cannot be removed by field redefinitions.\footnote{The deformation \eqref{firstV} is actually a generic one, whose existence we can prove for any higher spin algebra. Let us note, however, that nontrivial $\mathcal{V}_3(\omega,\omega,C)$ belongs to $HH^2(\hs,\mathcal{M})$, where $\mathcal{M}=\mathrm{Hom}(\hs,\hs)$ is endowed with a natural bimodule structure over $\hs$. Also, the matrix extension, mentioned in footnote \ref{matrixcomment}, makes the Chevalley-Eilenberg problem equivalent to the Hochschild one for matrices large enough. Therefore, we cannot exclude exceptional solutions that belong to $HH^2(\hs,\mathcal{M})$ and are not given by \eqref{firstV}, or are Chevalley-Eilenberg cocycles that are not obtainable from the Hochschild ones.} Note that if we drop $\pi$ in \eqref{twistedcocycle} we get a usual Hochschild two-cocycle that would induce a deformation of $\hs$ as an associative algebra (usually higher spin algebras are rigid and there are none). The observation is that while $\phi$ does not induce any deformation of $\hs$, it does so for its simple extension that we introduce below, thereby reducing the problem to that of associative algebras.

The second step is to eliminate any explicit $\pi$. We take a bigger algebra, called smash product algebra,\footnote{Its elements are pairs $(a, \pi)$, $a\in hs$, $\pi\in \Gamma$ and the product is $(a,\pi)\circ (a',\pi')=(a\pi(a'),\pi\pi')$,  where $\pi(a')$ is the action of the automorphism $\pi$ on $\hs$.} $\hsdouble=\hs\rtimes \Gamma$, with $\Gamma$ being a finite group of automorphisms of $\hs$. In our (the simplest) case,  $\Gamma=\mathbb{Z}_2=\{\mathbf{1},\pi\}$ and, abusing notation, we can write elements of $\hsdouble$ as $a=a'\mathbf{1}+a''\pi$, $a', a''\in \hs$. Next, we enlarge the set of fields as to make $\omega$, $C$ be elements of $\hsdouble$. This trick allows us to get rid of any explicit automorphisms:
\besubeqs\label{problemgen}
\begin{align}
\dd\omega&=\omega\star \omega + \mathcal{V}_3(\omega,\omega,C)+\mathcal{V}_4(\omega,\omega,C,C)+O(C^3)=F^\omega(\omega,\omega|C)\,,\\
\dd C&=\omega\star C-C\star \omega+\mathcal{V}_3(\omega,C,C)+O(C^3)=F^C(\omega|C) \,.
\end{align}
\esubeqs
The truncation back to the initial field content corresponds to $\omega=\omega'\,\mathbf{1}$, $C=C''\pi$. We can also pack the fields into a single `string field' $\Phi=\omega +\epsilon C$, where $\epsilon^2=0$ is a formal ghost, as to write a much more concise
\begin{align}
\dd\Phi&=\mathcal{V}_2(\Phi,\Phi) + \mathcal{V}_3(\Phi,\Phi,\Phi)+\mathcal{V}_4(\Phi,\Phi,\Phi,\Phi)+O(\Phi^5)=F(\Phi)\,,
\end{align}
where 
\begin{align}\label{VAinfinity}
    \mathcal{V}_2(\omega,\omega)&=\omega\star \omega\,, &
    \mathcal{V}_2(\omega,C)&=\omega\star C\,, &
    \mathcal{V}_2(C,\omega)&=-C\star \omega\,, &
    \mathcal{V}_2(C,C)&=0\,.
\end{align}
That \eqref{twistedcocycle} involves $\pi$ means that we deform $\hsdouble$ `in the direction of'  $\pi\in \hsdouble$, i.e., $\phi\, \pi$ is a Hochschild two-cocycle of the smash product algebra $\hsdouble$. For genuine higher spin algebras it is important to have $\pi$ in \eqref{freesystem} and, hence, in \eqref{twistedcocycle}. From the viewpoint of  deformation theory there is no conceptual difference between various elements of $HH^2(\ass,\ass)$: all of them induce infinitesimal deformations of the algebra $\ass$. It can further be shown that   for a large class of higher spin algebras the first-order deformation can always be extended to all higher orders \cite{Sharapov:2017yde,Sharapov:2017lxr,Sharapov:2018hnl,Sharapov:2018kjz}. Therefore, from now on we assume that we have an associative algebra $\ass=\hsdouble$ that can be deformed into a one-parameter family of algebras $\hsdeformed$ with a formal deformation parameter $u$. The associative product in $\hsdeformed$ is given by
\begin{align}\label{fullproduct}
    a\ast b&= a\star b+\sum_{k>0} \phi_k(a,b) u^k\equiv \mu(a,b)\,, && \phi_1\equiv \phi\,\pi\,.
\end{align}
As a consequence of associativity, the maps $\phi_k$ obey certain nonlinear relations, which will be important below. 

The main statement \cite{Sharapov:2018kjz} is that all the vertices $\mathcal{V}_n$ can now be constructed (up to equivalence) in terms of the  bilinear maps  $\phi_k$ above. For example, 
\begin{align}\label{ffour}
    \mathcal{V}_4(\omega,\omega,C,C)&=\phi_2(\omega,\omega)\star C\star C +\phi_1(\phi_1(\omega,\omega),C)\star C\,,
\end{align}
and, even more ambitious,
\begin{align}
\begin{aligned}
    \mathcal{V}_5&(\omega,\omega,C,C,C)=\phi_1(\phi_1(\phi_1(\omega,\omega),C),C)\star C+\phi_2(\phi_1(\omega,\omega),C)\star C\star C+\\&+\phi_1(\phi_2(\omega,\omega),C)\star C\star C+\phi_1(\phi_2(\omega,\omega)\star C,C)\star C+\phi_3(\omega,\omega)\star C\star C\star C\,.
\end{aligned}
\end{align}
Therefore, the `wild' nonlinear equations of Formal Higher Spin Gravities are completely determined by a one-parameter family of associative algebras $\hsdeformed$. The latter describes a deformation of a given higher spin algebra $\hs$ extended by the automorphism $\pi$, the deformation happening along $\pi$. Clearly, associative algebras are more tame and easier to construct than $L_\infty$-algebras underlying \eqref{problem}. In order to systematically describe the vertices $\mathcal{V}_n$ we have to refer to the concept of $A_\infty$-algebras.

%%%%%%%%%%%%%%%%%%%%%%%%%%%%%%%%%%%%%%%%%%%%%%%%%%%%%%%%%%%%%
\subsubsection{\texorpdfstring{Vertices from $A_\infty$-algebras}{Vertices from A-infinity Algebras}}
%%%%%%%%%%%%%%%%%%%%%%%%%%%%%%%%%%%%%%%%%%%%%%%%%%%%%%%%%%%%%
We first define strong homotopy associative algebras or $A_\infty$-algebras \cite{Stasheff,Kajiura:2003ax} along with certain higher operations on them and then  present the equations that generate the vertices $\mathcal{V}_n$. Let $V$ be a graded vector space and $W=\mathrm{Hom}(TV,V)$ denote the space of all multi-linear maps on $V$. The space $W$ inherits naturally the grading from $V$. The Gerstenhaber product $\circ$ is a non-associative product defined by 
\begin{align}\label{comp-prod}
    \begin{aligned}
    (f\circ g)&(a_1, a_2,\ldots, a_{m+n-1} )\\
     &=\sum_{i=0}^{n-1}(-1)^{|g|\sum_{j=1}^i|a_j|} f(a_1, \ldots, a_i, g(a_{i+1}, \ldots, a_{i+m}),\ldots , a_{m+n-1})
    \end{aligned}
\end{align}
for all homogeneous maps $f\in \mathrm{Hom}(T^nV,V)$, $g\in \mathrm{Hom}(T^mV,V)$ and the vectors $a_i\in V$. Here  $|g|$ and $|a_i|$ stand for the degrees of the map $g$ and of the element $a_i$, respectively. Using the Gerstenhaber product, one can 
define the Gerstenhaber bracket 
\begin{align}
   \gers{f,g}=f\circ g-(-1)^{|f||g|}g\circ f\,,
\end{align}
which is graded skew-symmetric and obeys the graded Jacobi identity:
\begin{align}\label{JI}
   \gers{f,g}&=-(-1)^{|f||g|}\gers{g,f} \,, &&\gers{\gers{f,g},h}=\gers{f,\gers{g,h}}-(-1)^{|f||g|}\gers{g,\gers{f,h}}\,.
\end{align}
The structure of an $A_\infty$-algebra on $V$ is defined by a degree-one element $m\in W$ satisfying the condition:\footnote{Usually, in the literature, one shifts the original degrees on $V$ by one, by going to its suspension $V[1]$, and then applies the definition to $V[1]$. We prefer to avoid this step and grade $V$ appropriately from the very beginning.}
\begin{align}\label{mm}
    \gers{m,m}&=0 \,.
\end{align}
Expanding $m$ into the sum of homogeneous multi-linear maps as  
$$m=m_1+m_2+m_3+\ldots \,, \qquad m_n\in \mathrm{Hom}(T^nV,V)\,,$$
and substituting back into (\ref{mm}), one gets an infinite collection of quadratic relations on $m_n$'s known as the Stasheff identities \cite{Stasheff}. 

We also need a simple generalization of the Gerstenhaber product, called braces \cite{Kadeishvili98,Getzler93, GV}. The brace operation substitutes given $k$ maps $g_1,\ldots, g_k$ as arguments into another multi-linear map $f$ yielding a new map 
\begin{align}\notag
    f\{g_1,\ldots ,g_k\}(a_1,\ldots)&=\sum \pm f(a_1,\ldots,a_{i_1},g_1(a_{i_1+1},\ldots),\ldots,g_2(\ldots),\ldots,g_k(\ldots),\ldots)\,.
\end{align}
The sign factor is a product of the sign factors for each $g_i$. The latter is given by $|g_i|$ times the sum of the degrees of all $a$'s that are to the left of $g_i$, i.e., $|g_k|\sum_{i=1}^{i=i_k} |a_i|$. This is the standard Koszul sign convention. The Gerstenhaber product is the brace operation with one argument, $f\circ g=f\{g\}$. The braces have a number of remarkable properties, which we will not present here referring the reader to \cite{GV}.  It follows from (\ref{mm}) and the Jacobi identity (\ref{JI}) that  the operator $D=\gers{m,\bullet}$ defines a differential of degree one in the space $W$. Indeed, 
$$
D^2f=\gers{m,\gers{m,f}}=\frac12\gers{\gers{m,m},f}=0\qquad \forall f\in W\,.
$$
So, it makes sense to speak about the cohomology of $D$. Furthermore, $D$ differentiates the two-brace $m\{\bullet,\bullet\}$ by the rule 
$$
D(m\{f,g\})=-m\{Df,g\}-(-1)^{|f|} m\{f,Dg\}\,.
$$
Among other things, this implies that the two-brace $m\{f,g\}$ of $D$-cocycles $f$ and $g$ is again a $D$-cocycle, see \cite{Sharapov:2018xxx} for more details.  In such a way $m\{\bullet,\bullet\}$ gives rise to a multiplication operation in $D$-cohomology. 

We now define an auxiliary family $\Ainf(t)$ of $A_\infty$-algebras, which will generate, in a while,  the desired vertices $\mathcal{V}_n$. As a vector space $\Ainf(t)$ consists of two components, $V=V_{-1}\oplus V_0$, each being isomorphic to $\hsdeformed$ (to become $\omega$ and $C$ later). At $t=0$ we set
\begin{align}
m_2(a,b)=a\ast b\equiv \mu(a,b)\,,\qquad m_2(a,v)=a\ast v\,,\qquad m_2(v,a)=-v\ast a\,, \qquad m_2(v,w)=0\,,\notag
\end{align}
where $a,b\in V_{-1}$ and $v,w\in V_{0}$. Here the product \eqref{fullproduct} on $\hsdeformed$ is used. At $u=0$ the structure of $\Ainf(0)$ manifests the initial data $\mathcal{V}_2$, \eqref{VAinfinity}, for the deformation. The higher structure maps in the expansion $m(t)=m_2+t m_3 +t^2m_4+\ldots$ can be generated by solving the evolution equation supplemented by two algebraic constraints that are consistent with the evolution:
\begin{align}\label{evolution}
   \pl_t m&= m\{m', \pl \}\,,&  \gers{m,\pl}&=0\,, & \gers{m,m}&=0\,,
\end{align}
where $m'=\pl_u m$ and $\pl$ is a somewhat trivial degree $(-1)$ differential that maps $V_0$ identically to $V_{-1}$ and annihilates $V_{-1}$. Essentially, this is a formal way to identify the module $V_{0}$ with the algebra $V_{-1}$. It is not hard to show \cite{Sharapov:2018xxx} that the system \eqref{evolution} is in involution. The explicit formula for all the vertices $\mathcal{V}_n$ is now obtained by restricting all $m_n$'s to the diagonal and setting $u=0$
\begin{align}
    \mathcal{V}_n(\omega, \omega, C,\ldots,C)&= m_n(\omega, \omega, C,\ldots,C)\Big|_{u=0}\,, \qquad \text{etc.}
\end{align}
The rule is that all arguments from $V_{-1}$ should be replaced by $\omega$, while the arguments from $V_0$ by $C$. 

Restriction to the diagonal effectively antisymmetrizes the arguments of the structure maps $m_n$'s producing an $L_\infty$-structure from an $A_\infty$ one. This antisymmetrization procedure is completely  analogous to 
the construction of the Lie bracket as the commutator in an associative algebra. A general lesson to learn  is that some of $L_\infty$-algebras are more easily deformed in terms of the underlying $A_\infty$-algebras.

Let us illustrate that the evolution equation (\ref{evolution}) reproduces the first few vertices. At the lowest order we need to find $m_3$ from $m_2\{ m_2',\pl\}$, with $m_2$ being given by the product $\mu$, \eqref{fullproduct}, in the deformed algebra $\hsdeformed$. We find
\begin{align*}
    m_3&= m_2\{ m_2', \pl \} && \longrightarrow && 
    \begin{cases}
        +\mu(\mu'(\omega,\omega), \pl(C))\stackrel{\scriptstyle{u=0}}{=}+\phi_1(\omega,\omega)\star C\,,\\
        +\mu(\mu'(\omega,C), \pl(C))\stackrel{\scriptstyle{u=0}}{=}+\phi_1(\omega,C)\star C\,,\\
        -\mu(\mu'(C,\omega), \pl(C))\stackrel{\scriptstyle{u=0}}{=}-\phi_1(C,\omega)\star C\,,
    \end{cases}
\end{align*}
where on the right we evaluated the map on the left for various triplets of arguments, all other orderings resulting in identically vanishing maps.\footnote{This is a particular choice, all other orderings are canonically equivalent, i.e., they are related by formal redefinitions in \cite{Sharapov:2018kjz}.} At the second order in $t$ we find
\begin{align}
   2 m_4&= m_3\{  m_2', \pl \}+m_2\{ m_3', \pl\}\,,
\end{align}
and hence, for example, 
\begin{align}
   m_4(\omega,\omega,C,C)&=\mu(\mu'(\mu'(\omega,\omega), \pl(C)), \pl(C))+\tfrac12\mu(\mu(\mu''(\omega,\omega), \pl(C)),\pl(C))\,,
\end{align}
which results in the following vertex:
\begin{align}
   \mathcal{V}_4(\omega,\omega,C,C)=&m_4(\omega,\omega,C,C)\Big|_{u=0}= \phi_1(\phi_1(\omega,\omega),C)\star C+ \phi_2(\omega,\omega)\star C\star C\,,
\end{align}
This is in agreement with \eqref{ffour}. Note that the evolution parameter $t$ behaves as a dimensionless coupling constant that accompanies expansion in powers of $C$. We conclude that the evolution equation \eqref{evolution} reconstructs the equations of the Formal Higher Spin Gravity associated to a given higher spin algebra $\hs$, whose deformation in the direction of $\pi$ is determined by $\hsdeformed$. We note that it is not necessary to set $u=0$ at the end of all calculations in order to get formally consistent equations. It is only needed to fulfill the boundary condition \eqref{VAinfinity} that the deformation starts out from the higher spin algebra. Finally, our construct allows one to convert any one-parameter family of algebras (not necessarily of higher spin origin) into highly nontrivial gauge-invariant equations.

%%%%%%%%%%%%%%%%%%%%%%%%%%%%%%%%%%%%%%%%%%%%%%%%%%%%%%%%%%%%%
\subsection{Solution Space and Lax Pair}
\label{subsec:lax}
%%%%%%%%%%%%%%%%%%%%%%%%%%%%%%%%%%%%%%%%%%%%%%%%%%%%%%%%%%%%%
We have constructed a Formal Higher Spin Gravity for any given higher spin algebra, provided that the deformation $\hsdeformed$ of $\hsdouble$ is known (proven to exist, at least). In practice, it is easy not only to prove that $\hsdeformed$ exists, but even to construct it. The equations we are interested in are obtained by constructing maps $m(t)$ according to \eqref{evolution} and setting $u=0$ at the end. It is interesting to look at  these equations at $u\neq 0$. The first few terms are\footnote{We use the short-hand notation $a\ast'b\equiv \pl_u \mu(a,b)$, $a\ast''b\equiv \pl_u^2 \mu(a,b)$.}
\besubeqs\label{usystem}
\begin{align}
    \dd\aomega&=\aomega \ast \aomega+ t(\aomega\ast'\aomega)\ast \auxC+t^2\tfrac12(\aomega\ast''\aomega)\ast \auxC\ast \auxC+t^2((\aomega\ast'\aomega)\ast' \auxC)\ast \auxC+\ldots\,,\\
    \dd \auxC&=\aomega\ast \auxC-\auxC\ast \aomega+t(\aomega\ast'\auxC)\ast \auxC-t(\auxC\ast' \aomega)\ast \auxC+\ldots\,,
\end{align}
\esubeqs
where the fields $\aomega\equiv \aomega(u;x)$, $\auxC\equiv \auxC(u;x)$ take values in $\hsdeformed$. 
The fact that the equations are completely determined by an associative algebra $\hsdeformed$ allows us to explicitly construct formal solutions in terms of a rather trivial system
\besubeqs\label{trivialsystem}
\begin{align}
    \dd\momega&=\momega \ast \momega\,, &&(\equiv \momega\star \momega+u\phi_1(\momega,\momega)+\ldots)\,,\\
    \dd \mC&=\momega\ast \mC-\mC\ast \momega\,, &&(\equiv [\momega,\mC]_\star+u\phi_1(\momega,\mC)-u\phi_1(\mC,\momega)+\ldots)\,,
\end{align}
\esubeqs
where $\momega\equiv \momega(u;x)$, $\mC\equiv \mC(u;x)$ take values in $\hsdeformed$. One can see that \eqref{usystem} is satisfied by 
\besubeqs\label{mauxmap}
\begin{align}
    \aomega&=\momega+t\momega' \ast \mC+t^2\tfrac12 \momega''\ast \mC\ast \mC+t^2\momega'\ast \mC'\ast \mC+t^2(\momega'\ast' \mC)\ast \mC+\ldots\,,\\ 
    \auxC&=\mC+t\mC'\ast \mC+\ldots\,,
\end{align}
\esubeqs
provided $\momega$ and $\mC$ obey \eqref{trivialsystem}. The fields $(\momega, \mC)$ satisfying \eqref{trivialsystem} may be thought of as defining a Lax pair for \eqref{usystem} with the identification between the fields given by \eqref{mauxmap}. Upon this interpretation the deformation parameter $u$ plays the role of the spectral parameter. In a topologically trivial situation, Eqs. \eqref{trivialsystem} can be solved in a pure gauge form, namely, $\momega=\mg^{-1}\ast \dd \mg$, $\mC=\mg^{-1}\ast \mC_0 \ast \mg$. Eqs. \eqref{usystem} are defined for all $u$ and setting $u=0$ gives back the equations of the Formal Higher Spin Gravity.

There is an apparent paradox here since we can construct solutions of a clearly nontrivial system of equations \eqref{usystem} and, hence, \eqref{problemgen} in terms of a trivial system \eqref{trivialsystem}. The resolution of the paradox is that the map (\ref{mauxmap}) involves the derivatives of fields with respect to the spectral parameter $u$. Therefore, we cannot express solutions of \eqref{usystem} at $u=0$ in terms of \eqref{trivialsystem} at some other fixed $u$. What is happening is that we are mapping the whole $u$-family of solutions to \eqref{trivialsystem} into solutions of \eqref{usystem}. Obviously, this is not a field-redefinition.\footnote{This might look similar to the `integration flow' discussed in \cite{Prokushkin:1998bq} for a particular realization of the $3d$ higher spin system, and in \cite{Didenko:2014dwa} for the $4d$ case. However, the idea is right the opposite. In \cite{Prokushkin:1998bq}, one maps solutions of one trivial system of type \eqref{trivialsystem} to another trivial system of type \eqref{trivialsystem}, the map taking place in a much bigger space of certain resolution of the higher spin algebra (the so-called $z$ variables). Here, we construct solutions of the nontrivial system \eqref{usystem} in terms of a one-parameter family of solutions to the trivial system \eqref{trivialsystem}.} 

After packing $\omega$ and $C$ into the string field $\Phi=\omega+\epsilon\, C$, the complete map between \eqref{trivialsystem} and \eqref{usystem} can be reconstructed by solving another evolution equation \cite{Sharapov:2018xxx}:
\begin{align}
    D_t\Phi=m\{D_u,\pl\}(\Phi)\,, && \big(\equiv \sum_k m\{D_u,\pl\}(\underbrace{\Phi,\ldots,\Phi}_k)\big)\,.
\end{align}
Here $D_t(\bullet)$ and $D_u(\bullet)$ are $t$ and $u$ derivatives understood as linear maps of degree-zero (this is useful in order to write very short $m\{D_u,\pl\}(\Phi)$ with all arguments hidden). The initial value at $t=0$ is $\Phi=\momega+\epsilon\, \mC$. At the first order we have
\begin{align}
    \Phi_1=m_2\{D_u,\pl\}(\Phi_0,\Phi_0)=
        \momega'\ast \mC+\epsilon\,\mC'\ast \mC\,,
\end{align}
which coincides with the order $\mathcal{O}(t)$ terms in \eqref{mauxmap}. At the second order we find
\begin{align}
\begin{aligned}
    2\Phi_2&=m_3\{D_u,\pl\}(\Phi_0,\Phi_0,\Phi_0)+m_2\{D_u,\pl\}(\Phi_1,\Phi_0)+m_2\{D_u,\pl\}(\Phi_0,\Phi_1)=\\
    &=2(\momega'\ast' \mC)\ast \mC+2 \momega'\ast\mC'\ast\mC+\momega''\ast \mC\ast \mC+\epsilon(\mC''\ast \mC\ast \mC+\ldots)\,,
\end{aligned}
\end{align}
which coincides with the order $\mathcal{O}(t^2)$ terms in \eqref{mauxmap}. To summarize, the equations of motion of the Formal Higher Spin Gravity are equivalent to the Lax equation \eqref{trivialsystem}. This also allows us to write down the most general solution of the equations.

%%%%%%%%%%%%%%%%%%%%%%%%%%%%%%%%%%%%%%%%%%%%%%%%%%%%%%%%%%%%%
\subsection{Higher Spin Waves}
\label{subsec:HSwaves}
%%%%%%%%%%%%%%%%%%%%%%%%%%%%%%%%%%%%%%%%%%%%%%%%%%%%%%%%%%%%%
The results above have a simple spin-off that shows that appropriate multiplets of higher spin fields can consistently propagate over much more general backgrounds than just maximally symmetric spaces (flat space and (anti)-de Sitter spaces). Indeed, it has long been thought that massless higher spin fields do not propagate on general gravitational backgrounds. A classical argument of \cite{Aragone:1979hx} is that upon replacing $\partial_\mu$ in the Fronsdal equations with the covariant derivative $\nabla_\mu$ associated with a generic metric $g_{\mu\nu}$, one finds the full four-index Riemann tensor $R_{\mu\nu,\lambda\rho}$ as an obstruction to gauge invariance. This does not happen for $s\leq 2$ and is the simplest sign of the numerous troubles that usually accompany higher spin ($s>2$) massless fields. 

It is clear that the most symmetric backgrounds in any higher-spin theory are given by flat connections of a given higher spin algebra $\hs$:
\begin{align}
    d\Omega&= \Omega\star \Omega\,.
\end{align}
The global symmetries of such backgrounds are isomorphic to the higher spin algebra $\hs$. Anti-de Sitter space is just the simplest representative of the maximally-symmetric backgrounds (in the higher spin sense). The linearized equations 
\besubeqs\label{HSwaves}
\begin{align}
    d\omega&= \Omega\star \omega+\omega\star \Omega +\phi(\Omega,\Omega)\star \pi(C)\,,\\
    dC&=\Omega\star C-C\star \pi(\Omega)\,,
\end{align}
\esubeqs
are consistent thanks to $\phi$ being a two-cocycle of the higher spin algebra \eqref{twistedcocycle}. The equations are invariant under the gauge transformations
\begin{align}
    \delta_\xi \omega&=d\xi -[\Omega,\xi]_\star\,,  &\delta_\xi C&=0\,,
\end{align}
so that  $C$ are gauge invariant curvatures. The global symmetry parameters $\epsilon_0$ obey   
\begin{align}
    d\epsilon_0&= \Omega\star \epsilon_0-\epsilon_0\star \Omega\,.
\end{align}
When $\Omega$ describes $AdS_{d+1}$ (higher spin components are turned off), $\epsilon_0$ is a set of Killing tensors. Global symmetries act as
\besubeqs
\begin{align}
    \delta_0 \omega&= \epsilon_0 \star \omega-\omega \star \epsilon_0 +\phi_1(\epsilon_0,\Omega)\star \pi(C)-\phi_1(\Omega,\epsilon_0)\star \pi(C)\,, \\
    \delta_0C&= \epsilon_0 \star C-C \star \pi(\epsilon_0)\,.
\end{align}
\esubeqs
Note that the term with the Hochschild cocycle is present even for the anti-de Sitter background. Equations \eqref{HSwaves} for higher spin waves can also be solved with the help of Sect. \ref{subsec:lax}. To summarize, higher spin fields can consistently propagate on higher spin flat backgrounds \cite{Sharapov:2017yde,Bekaert:2017bpy,Grigoriev:2018wrx}. This result illustrates the physical meaning of the, otherwise quite abstract, Hochschild two-cocycle. Another appealing feature of Eqs. \eqref{HSwaves} is that they should not be affected by the nonlocalities of higher spin interactions.

%%%%%%%%%%%%%%%%%%%%%%%%%%%%%%%%%%%%%%%%%%%%%%%%%%%%%%%%%%%%%
\subsection{Observables}
%%%%%%%%%%%%%%%%%%%%%%%%%%%%%%%%%%%%%%%%%%%%%%%%%%%%%%%%%%%%%
A general expectation is that the symmetries of higher spin theories are rich enough as to fix any meaningful physical observable. If this is true, then the nonlocality problem can, to large extent, be avoided. An important set of such observables are holographic correlation functions. Indeed, it is known \cite{Maldacena:2011jn,Boulanger:2013zza,Alba:2013yda,Alba:2015upa} that unbroken higher spin symmetries, i.e., higher spin algebra itself, fix all holographic correlation functions and imply that they are those of a free CFT (obviously, of the same free CFT that determines the higher spin algebra). This could be the end of the story if one would be able to show that boundary conditions preserving the full amount of higher spin symmetry can be imposed and are not destroyed by the nonlocality of interactions and by quantum corrections. Alternative and mixed boundary conditions are also sometimes possible \cite{Klebanov:1999tb,Witten:2001ua,Chang:2012kt} and should inherit a part of the nontrivial structure of the deformed higher spin symmetry \cite{Maldacena:2012sf,Sharapov:2018kjz}. In any case, it makes sense to ask what are the quantities, observables, at our disposal that behave nicely (invariant or covariant) under the deformed higher spin symmetries.   

We assume that all observables have a smooth limit when interactions are removed and everything is reduced to a higher spin algebra $\hs$. Therefore, the first general question is about the cohomology of higher spin algebras. The next question is whether the relevant cohomology groups deform smoothly when interactions are turned on. The observation of Sect. \ref{sec:FHSG} allows us to reduce these questions to the cohomology of the deformed higher spin algebra $\hsdeformed$ (since the information hidden in the full equations is the same). This is a great simplification: instead of classifying invariants of the nonlinear equations of motion, we can study the cohomology of the deformed higher spin algebra $\hsdeformed$. 

There are two nontrivial cocycles that any higher spin algebra $\hs$ comes equipped with. The two-cocycle $\phi$ in the twisted representation, \eqref{twistedcocycle}, and the trace $\mathrm{Tr}$.\footnote{Trace can be viewed as a zero-cocycle in the module dual to the adjoint one.}${}^{,}$\footnote{Whenever a given higher spin algebra admits a realization via the Weyl algebra, one can use the results of \cite{AFLS}. In general, the cohomology of the Weyl algebra is very sparse. } The two-cocycle is a necessary prerequisite  for higher spin gravities to exist. The existence of the trace is inherited from the universal enveloping $U(so(d,2))$ realization. There are two natural operations on cohomology: the cup product and the Gerstenhaber bracket.\footnote{We do not discuss higher operations like Massey products.} Since the associativity of $\hsdeformed$ implies $\gers{\phi_1,\phi_1}=\delta\phi_2$, the Gerstenhaber bracket does not give anything new, cf. (\ref{twistedcocycle}).  The cup product generates the sequence of higher cocycles $\phi^{(k)}=\phi_1\cup\phi_1\cup \cdots \cup \phi_1$. Remembering that the original cocycle is in the twisted-adjoint representation $\phi_1=\phi \pi$, we see, for example, that there may exist a nontrivial  four-cocycle $\phi\cup \pi(\phi)$ with values in the adjoint representation. In the $AdS_4$ case, where the higher spin algebra is related to the Weyl algebra $A_2$, the four-cocycle above was found in \cite{FFS}. In general, we expect $\phi^{(k)}$ to be trivial for $k$ large enough, so that we have cocycles of degrees $2,4,\ldots,2[d/2]$. These cocycles can be uplifted to the cocycles of the deformed algebra $\hsdeformed$. Indeed, the $u$-derivative of the deformed product $a\ast b\equiv\mu(a,b)$, \eqref{fullproduct},
\begin{align}
    \phi_u(a,b)&=\pl_u \mu(a,b)
\end{align}
is a nontrivial Hochschild two-cocycle of $\hsdeformed$ reducing to $\phi_1$ at $u=0$.

The trace leads to a number of interesting invariants. First of all, there are scalar invariants given by the on-shell closed zero-forms\footnote{When the functional class includes generalized functions \cite{Sezgin:2011hq,Colombo:2012jx,Didenko:2013bj,Didenko:2012tv,Bonezzi:2017vha} one can map between adjoint and twisted-adjoint modules, which allows to extend the number of invariants even further.}
\begin{align}\label{Invzero}
    I_n(u)&=\mathrm{Tr}[\,\underbrace{\mC\ast \mC\ast \cdots \ast \mC}_{n}\,]\,.
\end{align}
These are the usual integrals of motion associated with any Lax pair. Zero-form invariants were discussed in a number of papers \cite{Sezgin:2011hq,Colombo:2012jx,Didenko:2013bj,Didenko:2012tv,Bonezzi:2017vha}, but those, in general, differ from the invariants above.\footnote{A natural idea has been \cite{Sezgin:2011hq,Colombo:2012jx} to compute invariants of type $\mathrm{tr}(B\star B\star ...\star B)$, where $B=B(Y,Z|x)$ is a master field of \cite{Vasiliev:1999ba}. Such invariants, when computed on the solutions of the free equations, are known to reproduce free CFT correlation functions \cite{Colombo:2012jx,Didenko:2013bj,Didenko:2012tv,Bonezzi:2017vha}. However, they lead to infinite results beyond the linear approximation \cite{Colombo:2012jx}. Our findings show that finite zero-form invariants do exist. The apparent paradox can be explained as follows. The master field $B$ takes values in some resolution of the Hochschild complex \cite{Sharapov:2017yde,Sharapov:2017lxr}. The resolution is well-suited to give an explicit formula for the Hochschild two-cocycle (and higher orders as well). The trace, however, is a different cohomology and the same resolution may not be appropriate to get it (the super-trace on the $Y,Z$-algebra does not descend to a well-defined trace that would be the one on the deformed higher spin algebra). It may well be that there is some regularization that still allows one to reproduce the right trace from the resolution.} The zero-form invariants are the candidates for the holographic correlation functions. Indeed, at $u=0$, i.e., in the free field limit, they were shown \cite{Colombo:2012jx,Didenko:2013bj,Didenko:2012tv,Bonezzi:2017vha} to give correlation functions of higher spin currents in the dual free CFT. 

Another set of invariants is given by the on-shell closed $(2n+1)$-forms\footnote{For $n=1$ at the lowest order, $J_1$ is the spin-one gauge potential since to the lowest order in $C$, $d\omega|_{s=1}=0$. }
\begin{align}
    J_{2n+1}&=\mathrm{Tr}[\,\underbrace{\momega\ast \momega\ast\cdots \ast\momega}_{2n+1}\,]\,,&& n=1,2,\ldots\,.
\end{align}
One can also combine cup products of $\phi_u$ (or $\phi_1=\phi_{u=0}$ for the $\hs$ approximation to $\hsdeformed$) with the trace (and take various orderings of $\phi_u$ and $\mC$):
\begin{align}
    J_{2n,m}&=\mathrm{Tr}[\,\underbrace{\phi_u(\momega,\momega)\ast \cdots\ast \phi_u(\momega,\momega)}_n\ast \underbrace{\mC \ast\mC\ast \cdots \ast \mC}_m\,]\,.
\end{align}
These are some of the local invariants (it seems to be a complete list though). Nonlocal invariants include Wilson lines (see \cite{Sezgin:2011hq} for a related discussion) and, possibly, other quantities that are gauge invariant up to a total derivative.

%%%%%%%%%%%%%%%%%%%%%%%%%%%%%%%%%%%%%%%%%%%%%%%%%%%%%%%%%%
\subsection{Comments}
\label{sec:}
%%%%%%%%%%%%%%%%%%%%%%%%%%%%%%%%%%%%%%%%%%%%%%%%%%%%%%%%%%
The main result is that a seemingly complicated problem of constructing formally consistent equations that gauge a given higher spin algebra $\hs$ boils down to a much simpler problem of deforming $\hs$ in the direction of the $\pi$-automorphism. The complexity of the $L_\infty$-algebra underlying the nonlinear field equations gets reduced to the usual deformation  problem for associative algebras. The latter deformation can be proven to exist for a large class of algebras \cite{Sharapov:2017yde,Sharapov:2017lxr,Sharapov:2018hnl,Sharapov:2018kjz}. In practice, any such deformation is very easy to construct. Our approach also allows one to avoid all additional structures introduced in \cite{Vasiliev:1999ba}, e.g. the $z$-variables and additional fields. More generally, given any one-parameter family of associative algebras we can write formally consistent and gauge invariant equations following Sect. \ref{sec:FHSG}. Let us briefly discuss some special cases that are already covered in the literature.  

\noindent {\it Type-A in arbitrary dimension.} The higher spin algebra $\hs_A$ is the symmetry algebra of the free conformal scalar field \cite{Eastwood:2002su}. There are several realizations of $\hs_A$ \cite{Eastwood:2002su,Vasiliev:2003ev,Fernando:2015tiu}. The realization\footnote{$\ga,\gb,\ldots=1,2$ are the $sp(2)$ indices. Our convention is that $\epsilon^{\ga\gb}\epsilon_{\ga\gc}=\delta^\gb_\gc$, $y^\ga=\epsilon^{\ga\gb}y_\gb$ and $y^\ga\epsilon_{\ga\gb}=y_\gb$. $a,b,\ldots = 0,...,d$ are indices of the Lorentz algebra $so(d,1)$.} of \cite{Vasiliev:2003ev} is as the Weyl algebra $A_{2(d+2)}=A_{2(d+1)}\otimes A_1$
\begin{align}
    [y^a_\ga, y^b_\gb]&=2i \eta^{ab}\epsilon_{\ga\gb}\,, & [y_\ga,y_\gb]&=-2i\epsilon_{\ga\gb}\,,
\end{align}
with an $sp(2)$-subalgebra gauged.  The $sp(2)$-generators are $t_{\ga\gb}=-\tfrac{i}{4}\{y^a_\ga, y_{a\gb}\}+\tfrac{i}{4}\{y_\ga,y_\gb\}$. The higher spin algebra $\hs_A$ is generated by the simplest $sp(2)$-invariants: the AdS translations  $P^a=\tfrac{i}4\{y^a_\ga,y^\ga\}$ and the Lorentz generators $L^{ab}=\tfrac{i}4\{y^a_\ga,y^{b\ga}\}$. The $\pi$-map can be taken simply as $\pi(y_\ga)=-y_\ga$.\footnote{One could also try $\pi(y_\ga^a)=-y_\ga^a$, but this realization cannot be deformed. } Therefore, the $\mathbb{Z}_2$-extension, which we need to deform, can be realized by adding one more generator $k$ such that  $k^2=1$, $\{k,y_\ga\}=0$, and $[k,y^a_\ga]=0$. Essentially, the $y^a_\ga$ oscillators do not participate in the deformation. The deformation of the smash-product algebra $A_1\rtimes \mathbb{Z}_2$ generated by $y_\ga$ and $k$ is well-known as the deformed oscillator algebra $Aq(u)$ \cite{Yang:1951pyq,Mukunda:1980fv,Vasiliev:1989re}:
\begin{align}
    [q_\ga,q_\gb]&=2i \epsilon_{\ga\gb}(1+u K)\,, &&\{q_\ga,K\}=0\,, && K^2=1\,.
\end{align}   
Therefore, one simply replaces $y_\ga$ by $q_\ga$ everywhere to get the deformed higher spin algebra, which completes the study. The equations of motion are generated by the algorithm of Sect. \ref{sec:FHSG} and should be equivalent to the ones that can be extracted from \cite{Vasiliev:2003ev} and \cite{Bekaert:2017bpy}.

\noindent {\it Four dimensions.} This is one of the most interesting cases. The simplest higher spin algebra is the even subalgebra of the Weyl algebra $A_2=A_1\otimes A_1$ \cite{Dirac:1963ta,Gunaydin:1981yq}. The group of automorphisms is given by the Klein group $\mathbb{Z}_2\times \mathbb{Z}_2$, see \cite{Sharapov:2018kjz} for more detail. Applying the K\"unneth formula  shows that the Hochschild cohomology is two-dimensional in this case. Hence, there is an additional deformation parameter, which is in accordance with the conjectured duality \cite{Giombi:2011kc} involving Chern-Simons Matter theories. The deformed higher spin algebra results from $\hsdeformed=Aq(u e^{i\theta})\otimes Aq( ue^{-i\theta})$, where $u$ and $\theta$ are related to the microscopical parameters, the number of fields $N$ and the Chern-Simons level $k$. The large-$N$ correlation functions should be given by \eqref{Invzero}, \cite{Sharapov:2018kjz}. The equations associated with $\hsdeformed$ should be equivalent to the ones extractable from \cite{Vasiliev:1990vu}.

\noindent {\it Three dimensions.} Unfortunately, this is the case, where formal deformations do not seem to capture much of the dynamics. Higher spin fields do not have physical degrees of freedom in $3d$. The higher spin algebra is $\hs(\lambda)\oplus \hs(\lambda)$, see e.g. \cite{Gaberdiel:2012uj}. The zero-form $C$ is in the bi-fundamental of $\hs(\lambda)$ (instead of the twisted-adjoint) and describes a scalar field \cite{Prokushkin:1998bq}. It is clear from the field theory point of view that there are no nontrivial deformations that are linear in $C$ (no mixing between higher spin fields and the scalar at the free level) and quadratic in $C$ (all stress-tensors are formally exact \cite{Prokushkin:1999xq,Boulanger:2015ova,Skvortsov:2015lja}). Mathematically, the even subalgebra $A_1^e$ of the simplest Weyl algebra $A_1$ is a particular case of $\hs(\lambda)$ for $\lambda=\tfrac12$ and $A_1$ is known to have no relevant cohomology. Therefore, the complete system of equations is just 
\begin{align}\label{threed}
    dA&=A\ast A\,, & dB&=B\ast B\,, &dC&=A\ast C-C\ast B\,,
\end{align}
where $\ast$ is the product in $\hs(\lambda)$. The only possible formal deformation of this system is the shift of $\lambda$, which has nothing to do with interactions. It is this deformation that is described in \cite{Prokushkin:1998bq} (up to some decoration by matrix factors). As is clear from the minimal model holography \cite{Gaberdiel:2012uj}, the r.h.s. of the field equations should be altered by various interactions, e.g. by the stress-tensor of the scalar field. The simplest current interactions were added in \cite{Kessel:2015kna} using the compatibility of the global symmetries of \eqref{threed} with the deformations of the gauge symmetries induced by the usual Noether interactions.

Lastly, let us point out that all the deformations discussed above are based on particular realizations of various higher spin algebras as (subquotients) of the Weyl algebra (oscillators). The $\pi$-automorphism happens to be realized as the sign flip of one pair of the oscillators,  $\pi(y_\ga)=-y_\ga$, leaving the other oscillators and matrix factors intact (if any). This means that $\hs=A_1\otimes B$ (possibly modulo some relations), where $B$ is some associative algebra inert to the $\pi$-map. As a consequence, the deformed algebra can be realized as $Aq(u)\otimes B$. This explains the appearance of the deformed oscillator algebra in \cite{Vasiliev:1999ba}. Mathematically, all equations of \cite{Vasiliev:1999ba} are based on one and the same resolution of the Hochschild complex \cite{Sharapov:2017yde,Sharapov:2017lxr}, the one that leads to $Aq(u)$. 

In this regard it is worth mentioning that the Type-A,B theories of \cite{Bekaert:2017bpy,Grigoriev:2018wrx} seem to have nothing to do with the deformed oscillators and are based on some other resolution of the Hochschild complex. Below we briefly discuss another realization of the Type-B theory. In order to break the vicious circle of the deformed oscillators we also construct two new systems in five dimensions that are based on a different deformation/realization of higher spin algebras.

%%%%%%%%%%%%%%%%%%%%%%%%%%%%%%%%%%%%%%%%%%%%%%%%%%%%%%%%%%
\section{Type-B}
\label{sec:TypeB}
%%%%%%%%%%%%%%%%%%%%%%%%%%%%%%%%%%%%%%%%%%%%%%%%%%%%%%%%%%
As one more application of the general construction described above, let us discuss the formal Type-B theory --- the dual of the free/critical fermion CFT in dimension $d$. It has recently been worked out in \cite{Grigoriev:2018wrx} by using completely different techniques.\footnote{The parent approach \cite{Barnich:2004cr} used in \cite{Grigoriev:2018wrx} is well-suited for the formulation of PDE's, while the present study illustrates the algebraic aspects of the problem. } Since the field theory aspects were already discussed in \cite{Grigoriev:2018wrx}, we will present the deformed algebra only. 

At present there are several realizations of the Type-B higher spin algebra $\hs_B$: (i) as a symmetry of the free Dirac equation \cite{Nikitin1991fer}; (ii) oscillator realization \cite{Vasiliev:2004cm}; (iii) quasi-conformal realization \cite{Fernando:2015tiu}; (iv) universal enveloping realization \cite{Boulanger:2011se}. While it has been shown in \cite{Sharapov:2018kjz} that the universal enveloping realization of any higher spin algebra admits the deformation we are looking for, the realization (ii) is the most amenable to an explicit construction of the deformation. $\hs_B$ can be embedded into the super-Weyl algebra $\mathcal{A}=A_{2(d+2)}\times Cl_{d+2}$ as a certain subquotient \cite{Vasiliev:2004cm}. It is convenient to define the generators of $\mathcal{A}$ as
\begin{align}
[Y^A_\ga,Y^B_\gb]&=2i\eta^{AB}\epsilon_{\ga\gb}\,, && \{\phi_A,\phi_B\}=2\eta_{AB}\,. && A,B,\ldots=0,...,d+1
\end{align}
The crux of the matter is the Howe dual pair $so(d,2)\oplus osp(1|2) \subset \mathcal{A}$, whose generators are
\besubeqs
\begin{gather}
T^{AB}  = +\frac{i}{4}\epsilon^{\alpha\beta}\{Y^A_\alpha, Y^B_\beta\}  +\frac{1}{4}[\phi^A, \phi^B]\,,\\
t_\alpha = \frac{1}{2} Y^A_\alpha \phi_A\,, \qquad\qquad
t_{\alpha\beta} = -i \{t_\alpha, t_\beta\} = -\frac{i}{4}\{Y_\ga^A, Y_\gb^B\}\eta_{AB}\,.
\end{gather}
\esubeqs
The commutation relations manifesting the statement above read
\besubeqs
\begin{gather}
    [T_{AB},T_{CD}]=T_{AD}\eta_{BC}+\text{3 more}\,,\\
    [t_{\ga\gb},t_{\gc\gd}]= t_{\ga\gd}\epsilon_{\gb\gc}+\text{3 more}\,, \qquad \qquad  [t_{\ga\gb},t_{\gc}]= t_{\ga}\epsilon_{\gb\gc}+t_{\gb}\epsilon_{\ga\gc}\,,\\
    [T_{AB}, t_{\ga}]=0\,,\qquad\qquad\qquad \qquad[T_{AB}, t_{\ga\gb}]=0\,.
\end{gather}
\esubeqs
$\hs_B$ is defined by gauging the $osp(1|2)$ subalgebra, which can consistently be done thanks to the fact that $osp(1|2)$ commutes to $so(d,2)$. In more detail, an element of $\hs_B$ is the function $f(Y,\phi)$ that commutes with $osp(1|2)$ and is defined modulo $osp(1|2)$ gauge transformations:
\begin{align}
    hs \ni f(Y,\phi)&: &&t_\ga\star \rho(f)=f\star t_\ga\,,
        &f\sim f+ g^{\ga\gb}\star t_{\ga\ga} + g^\ga \star t_\ga\,.
\end{align}
Here, the automorphism $\rho$ is defined as $\rho[f(Y^A,\,\phi^A)]=f(Y^A,-\phi^A)$. 
Alternatively, following \cite{Grigoriev:2018wrx}, one can take the super-commutator of $t_\ga$ instead of the $\rho$-twisted commutator, as above. Notice that the gauge parameters $g^{\ga\gb}$ and $g^\ga$  are functions of $Y$'s and $\phi$'s transforming by appropriate representations of $osp(1|2)$.

The $so(d,2)$-generators $T_{AB}$ can be split into the Lorentz and translations generators:
\begin{align}
    P^a&=+\frac{i}{4} \{y^a_\ga, y^\ga\}+\frac14 [\phi^a,\phi]\,, &
    L^{ab}&=+\frac{i}{4} \{y^a_\ga, y^{b\ga}\}+\frac14 [\phi^a,\phi^b]\,,
\end{align}
where we split $\phi^A=(\phi^a,\phi)$, $Y^A_\ga=(y^a_\ga,y_\ga)$. 
The $\pi$-automorphism needs to be defined as to flip the translations and leave all other generators intact:
\begin{align}
\pi&:  && \pi(P^a)=-P^a,\qquad \pi(L^{ab})=L^{ab},\qquad \pi(t_\ga)=t_\ga\qquad \pi(t_{\ga\gb})=t_{\ga\gb}\,.
\end{align}
There are several options to achieve that, e.g.
\besubeqs
\begin{align}
    \pi_1&: && \pi(y^a_\ga)=y^a_\ga\,, \qquad \pi(y_\ga)=-y_\ga\,, \qquad \pi(\phi^a)=+\phi^a\,, \qquad \pi(\phi)=-\phi\,,\\
    \pi_2&: && \pi(y^a_\ga)=y^a_\ga\,, \qquad \pi(y_\ga)=-y_\ga\,, \qquad \pi(\phi^a)=-\phi^a\,, \qquad \pi(\phi)=+\phi\,.
\end{align}
\esubeqs
As it has been already noted for the Type-A case, different implementations of the twist map in a particular realization of the algebra may result in different deformations or none at all. Let us denote a nontrivial element of the $\pi$-center of the Clifford algebra as $\Gamma$, $\Gamma^2=1$ (depending on dimension $d$ and realization $\pi_{1,2}$ of the twist there can be different options for $\Gamma$: $\phi$ or $\phi_0\cdots \phi_d$). By definition, it realizes the twist map on $\phi^A$:
$\Gamma \phi^A \Gamma= \pi(\phi^A)$. 
The $\mathbb{Z}_2$-extension of $\hs_B$ is defined as (the first two relations are just $[k, y^a_\ga]=0$, $\{k,y_\ga\}=0$)
\begin{align}
     ky^a_\ga k&= \pi(y^a_\ga)=y^a_\ga\,, &
    ky_\ga k&= \pi(y_\ga)=-y_\ga\,,
     & k \phi^A k &= \pi(\phi^A)\,,
\end{align}
where for the action on $\phi^A$ we have the two options $\pi_{1,2}$. The deformed algebra is defined as
\besubeqs
\begin{align}
    [y^a_\ga, y^b_\gb]&= +2i \epsilon_{\ga\gb}\eta^{ab}\,, & [q_\ga,q_\gb]&=-2i \epsilon_{\ga\gb}(1+u k \Gamma)\,, & k^2&=1\,,\\
    \{\phi^a,\phi^b\}&=2\eta^{ab}\,, & \{\phi,\phi\}&=-2\,, & \{q_\ga,k\}&=0\,.
\end{align}
\esubeqs
Some additional relations, which may not be obvious include
\begin{align}
    [y^a_\ga,\phi^A]=[k,\Gamma]=[k\Gamma, \phi^A]=[y^a_\ga, q_\ga]=[\Gamma,q_\ga]=[q_\ga,y^a_\ga]=[q_\ga, \phi^A]=[\Gamma,y^a_\ga]=0\,.
\end{align}
Next, we define the deformed $osp(1|2)$ generators
\begin{align}
    t_{\ga\gb}&=-\frac{i}{4}\{y^a_\ga,y_{a\gb}\}+\frac{i}{4}\{q_\ga,q_{\gb}\} = -i \{t_\alpha, t_\beta\}\,, &
    t_\alpha &= \frac{1}{2} y^a_\alpha \phi_a-\frac12 q_\ga \phi\,.
\end{align}
It is important that the $osp(1|2)$ relations are preserved. The deformed Lorentz and translations generators are a bit more complicated, which is the main difference with the Type-A case. The translations get additional corrections as compared to the naive replacement $y_\ga\rightarrow q_\ga$:
\begin{align}
    \Pd^a_u&=+\frac{i}{4} \{y^a_\ga, q^\ga\}+\frac14 [\phi^a,\phi]+\frac{u}2k\phi^a\,, &
    \Ld^{ab}&=+\frac{i}{4} \{y^a_\ga, y^{b\ga}\}+\frac14 [\phi^a,\phi^b]\,.
\end{align}
These generators are the simplest $osp(1|2)$-invariants. The new deformed translations have correct commutation relations with the Lorentz generators:
\begin{equation}
    [\Ld^{ab}, \Pd^c_u]=\eta^{bc} \Pd^a_u -\eta^{ac} \Pd^b_u\,.
\end{equation}
There is also another useful $osp(1|2)$-invariant 
\begin{equation}
  {\rm K}^{ab}_u=\frac{i}{4}\{y^a_\alpha,y^{b\alpha}\}\phi -(\Pd^a\phi^b-\Pd^b\phi^a) - \frac{u k}4[\phi^a,\phi^b]\,,\qquad \{t_\alpha, K^{ab}_u\}=0\,.
\end{equation}
The commutator of two translations can now be written as
\begin{equation}\label{deformedB}
    [\Pd^a_u,\Pd^b_u]={\rm L}^{ab}+uk {\rm K}^{ab}_u\,.
\end{equation}
The strange metamorphoses with the former $so(d,2)$-algebra generators could have been easily predicted from the structure of the free equations \cite{Grigoriev:2018wrx}. The Weyl tensor belongs to the $(2,2,1)$-tensor of $so(d,2)$ at $u=0$. Therefore, the commutator \eqref{deformedB}, being dual to the first component of the Hochschild cocycle, is consistent with the free spin-two equations of motion. The deformed $\hs_B$ algebra can be defined as the algebra of $osp(1|2)$-invariants as before. Applying the construction of Sect. \ref{sec:FHSG} to the $u$-deformation gives the desired equations for the Formal Type-B Higher Spin Gravity.

%%%%%%%%%%%%%%%%%%%%%%%%%%%%%%%%%%%%%%%%%%%%%%%%%%%%%%%%%%
\section{Five Dimensions}
\label{sec:FiveD}
%%%%%%%%%%%%%%%%%%%%%%%%%%%%%%%%%%%%%%%%%%%%%%%%%%%%%%%%%%
The case of five dimensions, $AdS_5$, is of particular interest, see e.g. \cite{Sezgin:2001yf}. One feature that we will take advantage of is that there is a one-parameter family of higher spin algebras $\hs_\lambda(sl_4)$ \cite{Fernando:2009fq,Boulanger:2011se, Manvelyan:2013oua, Joung:2014qya} interpolating between higher spin algebras of massless $4d$ conformal fields of various spins.  This was explicitly worked out in \cite{Joung:2014qya}, including the structure constants. We will apply the construction of Sect. \ref{sec:FHSG} to this $\lambda$-deformation.

%%%%%%%%%%%%%%%%%%%%%%%%%%%%%%%%%%%%%%%%%%%%%%%%%%%%%%%%%%
\subsection{Higher Spin Algebra}
%%%%%%%%%%%%%%%%%%%%%%%%%%%%%%%%%%%%%%%%%%%%%%%%%%%%%%%%%%
The family of higher spin algebras can be embedded into the Weyl algebra $A_{4}$, whose generators obeys the commutation relations 
\begin{align}
[a_A,b^B]&=i\, \delta\fdu{A}{B}\,, && A,B,\ldots=1,\ldots,4\,.
\end{align}
They are interpreted as (anti)-fundamental of $su(2,2)$, i.e., as $\boldsymbol{4}\oplus \boldsymbol{\bar 4}$. Ignoring the reality conditions, the $gl(4)$ generators 
\begin{align}
    T\fdu{A}{B}&=-\frac{i}{2}\{a_A,b^B\}\,, && [T\fdu{A}{B}, T\fdu{C}{D}]=\delta^A_D T\fdu{C}{B}-\delta^B_C T\fdu{A}{D}\,,
\end{align}
can be split into the central $u(1)$ element $N=T\fdu{C}{C}$ and the traceless $S\fdu{A}{B}$ generators of $sl(4)$. The one-parameter family of algebras is defined by gauging the $u(1)$. To be precise, 
\begin{align}
    f(a,b)\in \hs_\lambda(sl_4) &: & [N,f]&=0\,, && f\sim f+ (N-\lambda) g\,,
\end{align}
where the gauge parameter $g$ is also a $u(1)$-singlet, $[N,g]=0$. The $u(1)$-singlet constraint means that $f(a,b)$ can be decomposed into monomials with equal number of $a$'s and $b$'s
\begin{align}
    f(a,b)&= \sum_k f\fud{A(k)}{B(k)}a_A\cdots a_A \, b^B\cdots b^B\,.
\end{align}
Gauge symmetry with $g$ allows one to make the Taylor coefficients traceless. Connections of $\hs_\lambda$ look like they should describe an infinite multiplet of totally-symmetric massless fields with spins $s=1,2,3,\ldots$.

The Lorentz algebra is $sp(4)\sim so(4,1)$ and we denote its invariant tensor $C^{AB}$, which is then used to raise and lower indices. It is convenient to introduce $Y_1^A=(a^A+b^A)$ and $Y_2^A=-i(a^A-b^A)$ with the commutation relations
\begin{align}
[Y^A_i, Y^B_j]&=2i \delta_{ij}C^{AB}\,.
\end{align}
The $AdS_5$ Lorentz and translations (plus $u(1)$) generators are 
\begin{align}
L^{AB}&=-\frac{i}4\{Y_i^A,Y_j^B\}\delta^{ij}\,, & T^{AB}&=-\frac{i}4\{Y_i^A,Y_j^B\}\epsilon^{ij}\,.
\end{align}
The translations are given by the $sp(4)$-traceless part of $T_{AB}$:
\begin{align}
P_{AB}&=T_{AB}-\frac{1}{4} C_{AB}T\fdu{K}{K} \,.
\end{align}

%%%%%%%%%%%%%%%%%%%%%%%%%%%%%%%%%%%%%%%%%%%%%%%%%%%%%%%%%%
\subsection{Topological Higher Spin Fields}
%%%%%%%%%%%%%%%%%%%%%%%%%%%%%%%%%%%%%%%%%%%%%%%%%%%%%%%%%%
We can combine the general construction of Formal Higher Spin Gravities from Sect. \ref{sec:FHSG} with the one-parameter family of algebras $\hs_\lambda\equiv \hs_\lambda(sl_4)$. The general properties can be inferred just from the very fact of existence of this one-parameter family. We use only the $\lambda$-deformation and do not turn on the deformation along the $\pi$-direction. Therefore, this is not the deformation that leads to the usual Higher Spin Gravities. Still we get a nonlinear system with higher spin fields. 

The fields are given by the one-form $\omega$ and zero-form $K$, both taking values in $\hs_\lambda$. The construction of Sect. \ref{sec:FHSG} allows us to write down equations whose lowest order terms look as\besubeqs
\begin{align}\label{hsomega}
    d\omega&=\omega\star \omega + \phi_1(\omega,\omega)\star K+\ldots \,,\\
    dK&=\omega\star K-K\star \omega+ \phi_1(\omega,K)\star K-\phi_1(K,\omega)\star K+\ldots \,.\label{hsC}
\end{align}
\esubeqs
One possible physical interpretation is as follows. Let us take $\Omega$ to be a flat connection and reduce the system to its linear approximation, as in Sect. \ref{subsec:HSwaves},
\begin{align}\label{omehahslin}
    d\omega&=\{\Omega, \omega\}+\phi_1(\Omega,\Omega)\star K\,, &
    dK&=\Omega\star K-K\star \Omega\,.
\end{align}
If $\Omega$ is $so(4,2)$-connection, then the $K$-equation describes a set of $AdS_5$ Killing tensors (starting from constants). Indeed, the only nontrivial equations in the $K$ sector are
\begin{align}
    \nabla_{a_1} K_{a_2\cdots a_s}+\text{permutations}&=0\,,
\end{align}
where the totally symmetric traceless tensors above result from $K^{a_2\cdots a_s}P_{a_2}\cdots P_{a_s}$ components of $K(P_a,L_{ab})$. For a given flat $\Omega$ the $\omega$-equation \eqref{omehahslin} describes a fixed configuration of higher spin fields whose curvature is determined by the Killing tensors $K$. 

If we take $K=c$, where $c$ is just a constant, then the $K$-equation \eqref{hsC} is obviously satisfied, while the $\omega$-equation \eqref{hsomega}   
\begin{align}
    d\omega&=\omega\star \omega +\sum_{k>0} \phi_k(\omega,\omega) c^k=\omega \ast_c \omega 
\end{align}
describes a flat connection of $\hs_{\lambda+c}$ algebra, which is expanded over that of $\hs_\lambda$. In the general case we keep $K$ non-singlet and the system is more interesting. The higher spin fields do not propagate any degrees of freedom, but acquire certain nontrivial values that are parameterized by Killing tensors $K$. This theory is not obstructed by the locality problems of higher spin theories.

%%%%%%%%%%%%%%%%%%%%%%%%%%%%%%%%%%%%%%%%%%%%%%%%%%%%%%%%%%
\subsection{Quasi-topological Higher Spin Fields}
%%%%%%%%%%%%%%%%%%%%%%%%%%%%%%%%%%%%%%%%%%%%%%%%%%%%%%%%%%
Let us propose another system that is still heavily based on $\hs_\lambda$, but does have propagating degrees of freedom. We need to realize the $\pi$-automorphism and this requires a doubling of the algebra. The reason is that, in the dual CFT picture, the automorphism is realized as the inversion map  mixing the positive and negative chirality fields. Note that $\hs_\lambda$ is the symmetry algebra of one irreducible free field (the precise relation between $\lambda$ and helicity is not important right now).\footnote{We are grateful to Karapet Mkrtchyan for the very useful discussion around this point.} Therefore, the first step is to double the algebra, which can be done by adding an idempotent element $\phi$ that commutes to the oscillators:
\begin{align}
[Y^A_i, Y^B_j]&=2i \eta_{ij}C^{AB}\,, && \phi^2=1\,, && [\phi,Y^A_i]=0\,.
\end{align}
The projectors on the two isomorphic copies $\Pi_\pm f(Y,\phi)$ are obtained with the help of $\Pi_\pm=\tfrac12(1\pm \phi)$. The $u(1)$ generator is defined as before $N=-\tfrac18\{Y^A_i,Y^B_j\} \epsilon^{ij}C_{AB}$. We first restrict to $u(1)$ singlets. Next, we
we take the quotient with respect to the ideal $I$ generated by
\begin{align}
   \mathcal{J}_\lambda= N-\lambda \phi\,,
\end{align}
which for $\Pi_+f$ corresponds to $N-\lambda$ and for $\Pi_-f$ corresponds to $N+\lambda$. Therefore, the quotient algebra is isomorphic to $\hs_{+\lambda}\oplus \hs_{-\lambda}$, as required.  In order to incorporate the twist we add a new generator $k$ that obeys
\begin{align}
k^2=1\,,&&k(Y^A_i)k=\pi(Y^A_i)\,, && \{k,\phi\}=0\,.
\end{align}
Here $\pi(Y^A_i)=\tau\fdu{i}{j}Y^A_j$ and $\tau$ preserves $\delta^{ij}$ and flips the sign of $\epsilon^{ij}$, i.e., it is any element of $o(2)$ with $\det \tau=-1$. 
That the initial algebra is $\hs_{+\lambda}\oplus \hs_{-\lambda}$ makes the generator of the ideal well-defined in the presence of the $\pi$-map, $k \mathcal{J}_\lambda k=-\mathcal{J}_\lambda$. 
It was shown in \cite{Sharapov:2018kjz} that, on general grounds, such an algebra admits a deformation along the $\pi$-direction and belongs to a one-parameter family of algebras. Together with $\lambda$ we have a two-parameter\footnote{If we reinstall the $\hbar$ that controls the star-product, then we have a three-parameter family of algebras. } family of algebras $\hs_{\lambda,\nu}$. Applying the construction of Sect. \ref{sec:FHSG} to the parameter $\nu$ gives rise to the expected five-dimensional higher spin equations. 

Let us turn off the $\pi$-deformation and apply the machinery of Sect. \ref{sec:FHSG} to the $\lambda$-parameter only. Even with $\phi$ and $k$ the structure constants of the algebra are directly related to those of $\hs_\lambda$, given in \cite{Joung:2014qya}. The fields $\omega$, $C$ take values in the algebra of $u(1)$-singlet functions $f(Y^A_i,k,\phi)$ quotiented by the ideal generated by $\mathcal{J}_\lambda$. Thanks to the dependence on $k$ the field $C$ contains zero-forms both in adjoint and twisted-adjoint representations of the higher spin algebra. Therefore, even without turning on the $\pi$-deformation, the resulting system of equations describes propagating degrees of freedom.\footnote{We need to restrict to those discrete values of $\lambda$ that correspond to the $4d$ free conformal fields.} What is missing is the two-cocycle that glues the part of $C$ in the $\pi$-twisted module to $\omega$, while the two-cocycle in the adjoint representation is present thanks to the $\lambda$-deformation. This gives another example of a higher spin system in five dimensions.

%%%%%%%%%%%%%%%%%%%%%%%%%%%%%%%%%%%%%%%%%%%%%%%%%%%%%%%%%%%%%
\section*{Acknowledgments}
\label{sec:Aknowledgements}
%%%%%%%%%%%%%%%%%%%%%%%%%%%%%%%%%%%%%%%%%%%%%%%%%%%%%%%%%%%%%
We are grateful to Maxim Grigoriev and Karapet Mkrtchyan for the very useful discussions. The work of E.S. was supported by the Russian Science Foundation grant 18-72-10123 in association with the Lebedev Physical Institute.

\providecommand{\href}[2]{#2}\begingroup\raggedright\endgroup

\end{document}